\documentclass{elsart_2}

\usepackage{graphicx}
\usepackage{natbib}
\bibpunct{(}{)}{;}{a}{}{,}

\begin{document}
\begin{frontmatter}

\title{Galactic globular clusters contribution to microlensing events?}

\author{Fabiana De Luca and Philippe Jetzer}

\address{
Institute for Theoretical Physics,  University of Z\"urich, 
Winterthurerstrasse 190, CH-8057 Z\"urich, Switzerland
}


\begin{abstract}
In this note we perform an analysis of the large set of microlensing events detected
so far toward the Galactic center with the purpose of investigating whether some 
of the dark lenses are located in Galactic globular clusters.
We find that in four cases some events might indeed be due to lenses
located in the globular clusters themselves.
We also give a rough estimate for the average lens mass
of the subset of events being highly aligned with Galactic globular cluster centers
and find that, under reasonable assumptions, 
the deflectors could most probably be either brown dwarfs, M-stars or stellar remnants.
\end{abstract}

\end{frontmatter}

\section{Introduction} 
\label{sec:INTRO}
\thispagestyle{plain}
Microlensing event analyses have been originally proposed as a tool to 
detect dark matter in form of MACHOs (\emph{MAssive Compact Halo Objects}) \citep{pacz86}. 
A remarkable number of collaborations has put considerable effort in the search
for such events toward various targets
and a large amount of detections has been collected.
Since the lens mass and velocity distributions are tightly related
with the spatial and time distributions of the events, the latter can 
be a useful tool to understand the structure, dynamics and 
the initial mass function in the low mass range of our own galaxy, as shown for instance
in \cite{wood} and \cite{calchi07} for the Milky Way.\\

Globular cluster could contain a sizeable amount of dark matter in form of
brown dwarfs or low mass stars. This is still an open issue and a possible way to test 
this is to use microlensing observations as suggested by \cite{pacz94}. The idea is to monitor
globular clusters in front of rich background of either the SMC, in the case of 47 Tuc,
or the galactic bulge. In this case, when the lens belongs to the cluster population,
its distance and velocity are roughly known. This way it is possible to get 
a more accurate estimate for the lens mass. 
Such a study has already been performed by \cite{straessle} for the globular clusters
in the direction of the galactic bulge region. Indeed, some events were found
which might be associated with lenses in globular clusters (\cite{straessle,sahu}). 
However, given the few events at disposal it has not been
possible to draw firm conclusions. In the meantime the number of microlensing events
detected toward the galactic bulge region has considerably increased and it is thus
reasonable to repeat that analysis.
We do not consider here the possibility that the source lies as well in the globular cluster,
as its probability is much smaller, nonetheless such suggestions
have been also studied (see also \cite{gyuk,malhotra,cardone}).
 
In this note we analyse the possible MACHO content in  
a large set of Galactic Globular Clusters (hereafter GGCs) some of which are highly aligned 
with a non negligible number of microlensing events 
detected toward the Galactic Center (hereafter GC).\\
The basic consideration is that, 
since the observed event duration (Einstein time, $t_E$)
is a function of the lens mass $t_E\sim m^{1/2}$,
an estimation of the latter can be drawn 
through reasonable assumptions on 
the spatial and velocity distributions of the lens and source populations.\\

Our aim, following the previous analysis in \cite{straessle},
is to get more stringent conclusions 
by enlarging the analysed data set
that now includes 4697 microlensing events
detected in the last years by the MACHO \citep{thomas}, EROS \citep{eros2}, 
OGLE (\cite{ogle}, $http://www.astrouw.edu.pl/\sim ogle/$),
 and MOA ($http://www.phys.canterbury.ac.nz/moa/$)
collaborations in direction of the GC.

This note is structured as follows.
An overview of the adopted models 
for the Galactic luminous components 
and GGCs is given in Sect. \ref{sec:models},
the results are discussed in Sect. \ref{sec:results} 
and Sect. \ref{sec:conclusions} is devoted to the conclusions.

\section{Models} \label{sec:models}
In our analysis we focus on the configuration in which the lens is hosted in a GGC
and the source is located either in the
Galactic disc or bulge (\cite{taillet95}, \cite{taillet96}).
The possibility that the source belongs to a GGC is neglected because GGC centers are very crowded regions, with at most $10^5$ stars, often located toward the GC, this making very unlikely the detection of microlensing of a GGC star, since very high resolutions and long observation periods (due to the very small optical depth) are required to provide valueable results.

As regards the GGCs,
the mass density of their luminous component 
is well described by a King function \citep{king}
\begin{equation}
\rho(r) = \frac{\rho(0)}{z^2}\big[\frac{\arccos[z]}{z}-\sqrt{1-z^2}\big], ~~z = \sqrt{\frac{r_c^2+r^2}{r_c^2+r_t^2}}, 
\label{density}
\end{equation}

that corresponds to a spherically symmetric mass distribution
whose surface density decreases with distance from the center
and rapidly drops to zero from $r=r_t$ on, as supported by star counts 
(for more details see \cite{binney,straessle}).\\
In eq. (\ref{density}) $r_c$ and $r_t$ are the GGC core and tidal radius, 
respectively, whose values are as given in \cite{harris},
while $\rho(0)$ has been calculated from the total GGC mass 
given in \cite{mandushev}.
In particular, out of the initial 150 GGCs given in \cite{harris}, 
we will only focus on a subset of 135 clusters for which the crossing 
of the two catalogs provides the complete set of necessary parameters.
\\
Since the MACHO distribution in GGCs is not known,
we assume for the dark matter the same mass density profile as given in eq. (\ref{density})
but rescaled by a factor 
$f=\frac{M_{dark}}{M_{tot}}$
which is the fraction of the total MACHO mass on the overall GGC mass,
that can even get the remarkable value of $1/2$ \citep{heggie}.\\
For the Galactic bulge we assume a density profile, $\rho_b$, as found
in \cite{stanek}, where the model that best fits
the observations suggests a triaxial, boxy shaped bulge with
$\rho_b = \rho_{0,b} exp[-r]$, $r = \sqrt{(x/x_o)^2+(y/y_o)^2+(z/z_o)^2}$
and major semiaxes $\{x_o,y_o,z_o\}=\{0.897,0.387,0.250\}\ pc$, 
the major axis being clockwise rotated on the Galactic plane
of an angle $\alpha=23.8^{\circ}$ with respect to the direction Sun-GC \citep{calchi07}.\\
Finally, the Galactic disc density profile is assumed to be exponentially decreasing
in both the Galactic plane and vertical direction with
$\rho_d = \rho_{0,d} exp\big[-\frac{R-R_{\odot}}{H}\big] sech^2\big[\frac{z}{h}\big]$, 
where $\{R,z\}$ are the distance from the GC on the equatorial plane and the height
above it, respectively, $R_{\odot}=8$ kpc is the distance   
of the Sun from the GC and $\{H,h\}=\{2.75,0.250\}$ kpc.
Our assumption on the Galactic disc structure
closely follows the model adopted by \cite{han&gould}
except for the fact that we neglect the contribution of a thick disc 
since this component should only provide minor contribution to the overall disc density
\citep{vallenari}. For further details on the bulge and disc density functions we refer to \cite{calchi07}.

\section{Results} \label{sec:results}
\begin{figure}[here]
\begin{center}
\includegraphics[width=8.5cm, height=8.5cm]{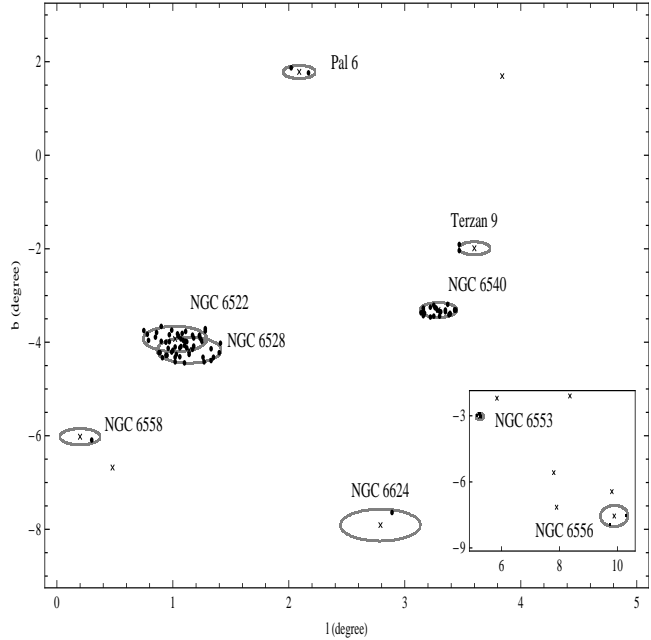}
\caption{ The points mark the MACHO, EROS, OGLE and MOA events detected behind GGCs. 
The crosses denote the centers of the GGCs.
Solid lines: contours on the sky plane of the sphere of radius equal to the cut radius of the GGC.
}
\label{fig:gcs}
\end{center}
\end{figure}

Aiming at discriminating among events due to lenses hosted either in GGCs
or in the Galactic bulge/disc, 
we first make a rough selection of events being aligned with a GGC.
In particular, for every given GGC, we consider a sphere of radius $r_t$,
 centered at the GGC center,
and we select, as a first step, only the events being included in one such contour, as shown in Fig. \ref{fig:gcs}.
By doing so, out of the original 4697 events, we are left with only 118.

The GGCs aligned with at least one event are given in Table \ref{tab:GGC1}.
Some properties are also reported, such as the tidal and the core radius, the distance 
from the Sun \citep{harris} and the number of included events.
Moreover we give the optical depth, $\tau$, \citep{k&p,calchi07}
\begin{equation}
\tau= \frac{\frac{4\pi G}{c^2}}{\int_{0}^{\infty}\rho_s dD_{os}}  \int_{D_{os}}^{\infty}\rho_s dD_{os} \int_0^{D_{os}} \rho_l D_{ol} \large(1-\frac{D_{ol}}{D_{os}}\large)dD_{ol} 
\end{equation}
calculated toward the GGC centers for
sources located in the Galactic bulge/disc and
lenses belonging to the GGC.
Here $D_{ol}$ ($D_{os}$) is the
observer-lens (observer-source) distance as measured along the 
line of sight observer-source (hereafter l.o.s.).
The different core radii and central densities explain
the large variations of the optical depth among the GCCs.
It is worth to underline that 
for both lenses and sources belonging to a GGC, the optical depth
is 2-3 orders of magnitude smaller than the reported $\tau$,
while the introduction of a thick disc (see \cite{calchi07} for a modelisation)
only implies variations of the order of $5\%$ or less.

\begin{table*}
\caption{GGCs being aligned with at least one detected microlensing event.
For each of them the number of aligned events $N_{tot}$ and
the corresponding average duration $<t_E>$ (in days) is given.
For every GGC, $r_t$ is the tidal radius (in pc), 
$r_c$ is the core radius (in pc), 
$r_{sun}$ is the distance of its center from the Sun (in kpc)
and $\tau$ is the optical depth toward its center
in units of $f\times 10^{-5}$, $f$ being the fraction 
of dark matter mass in the GGC (see section \ref{sec:models} for details). 
}
\label{tab:GGC1} 
\centering      
\begin{tabular}{c|cccccc}
\hline             
{Cluster ID} &$r_{sun}$&$r_t$& $r_c$&  $\tau$ & $N_{tot}$& $<t_E>$  \\
\hline                                                 
 Pal 6        & 5.9  &     14.3  & 1.13 & 0.24   &   2    & 76.3    \\
Terzan 9      & 6.5  &     15.5  & 0.06 & 0.78   &   2    & 17.8   \\
NGC 6522      & 7.8  &     37.3  & 0.11 & 6.06   &   36   & 16.8   \\
NGC 6528      & 7.9  &     38.1  & 0.21 & 1.01   &   38   & 25.2   \\
NGC 6540      & 3.7  &     10.2  & 0.03 & 23.74  &   29   & 22.8   \\
NGC 6553      & 6.0  &     14.2  & 0.96 & 1.01   &   7    & 43.0   \\
NGC 6558      & 7.4  &     22.5  & 0.06 & 7.20   &   1    & 24.5   \\
NGC 6624      & 7.9  &     47.2  & 0.14 & 4.43   &   1    & 223.0   \\
NGC 6656      & 3.2  &     27.0  & 1.32 & 0.75   &   2    & 112.7   \\
\hline
\end{tabular}
\end{table*}

\begin{table*}
\caption{GGCs including at least one inner event.
For each of them $N_{in}$ is the number of events inside a projected radius 
$r=2\times r_t/5$
and, for this subset of aligned events,
$<t_E>$ is the mean Einstein time (in days)
and $<m>$ is the average predicted lens mass in units of
solar masses.
$N_{GGC}$ ($N_{BD}$) is the number of events, out of $N_{in}$, that we expect to be due to GGC (Galactic bulge/disc) lenses.
$\Gamma_{exp}$ is the expected event rate in units of $f\times \mu_o^{-1/2}\times 10^{-3}/year$ (see text for details) while 
$n_{GGC}$ is $N_{GGC}$ per unit area (in $\mathrm{degree^{-2}}$).
}
\label{tab:GGC2} 
\centering      
\begin{tabular}{c|ccccccc}
\hline             
{Cluster ID} &$N_{in}$& $<t_E>$  & $<m>$  &$N_{BD}$    & $N_{GGC}$ &$\Gamma_{exp}$ &$n_{GGC}$\\
\hline                                                          
 NGC 6522     &   8   &   13.1   &  1.63  & 4.1$\pm$ 2.0& 3.9      &  0.66    & 51.4     \\
 NGC 6528     &   7   &   13.0   &  2.98  & 4.9$\pm$ 2.2& 2.1      &  0.09    & 27.5     \\
 NGC 6540     &   7   &   17.2   &  0.06  & 4.2$\pm$ 2.0& 2.8      &  1.56   & 112.3    \\
 NGC 6553     &   4   &   35.7   &  0.62  & 0.6$\pm$ 0.8& 3.4      &  0.08    & 185.4    \\

\hline

\end{tabular}
\end{table*}

Due to the GGC structure, we expect the predicted number of events to
be the largest toward
their centers and to decrease as we move toward their borders.
Since the alignment between an event and a projected cluster contour 
does not assure that the deflector belongs to
the GGC, this alignment possibly being accidental,
we make a further, rough selection
and consider only the events being included in the projected 
contour of a sphere centered at a GGC center and of radius $r_i=2\times r_t/5$
(this including on average $90 \%$ of the total cluster mass).
We then distinguish between $inner$ and $outer$ events, the former being inside $r_i$
and the latter being included in the circular ring of internal radius $r_i$ and outer radius $r_t$.
By doing so, we assume all the outer events to be due to Galactic 
bulge/disc deflectors (this possibly underestimating the events due to GGC lenses),
whereas we leave open the possibility that among the inner events
some could still be attributed to bulge/disc deflectors.
Object of our analysis is the subset of 28 inner events that we are left with,
among which 7 (17/4) have been detected by the MACHO (OGLE/MOA) collaboration.
Notice that, as shown in Fig. \ref{fig:gcs}, the projections of NGC 6522 and NGC 6528
superimpose and have some aligned events in common.
The latter are nevertheless far from the centers of both GGCs and are 
not included in our analysis.

An estimation of the predicted 
number of events, $N_{GGC}$, due to MACHOs in a given GGC,
can be roughly made as follows.
Assuming that all the outer events are due to Galactic bulge/disc lenses,
we calculate how many such events, $N_{BD}$, are expected in the inner region of a 
GGC contour assuming that the number of events 
is proportional to the covered area
and that the background source distribution is uniform inside every GGC contour.
Thus we assume that the microlensing rate for Galactic bulge/disc events
is constant over the entire small
area within the tidal radius of the considered globular cluster.
By doing so, $N_{BD}$ is simply proportional
to the monitored area. Clearly, also with these assumptions, which are reasonable,
 given the very small area considered, one expects fluctuations in the number
of events in a given area. We assume the
fluctuations to follow Poisson statistics, in which case they are given by
$\sim \sqrt{N_{BD}}$.
By doing so, for every GGC considered,
$N_{GGC}$ turns out to be around 2-4 per cluster (see Table \ref{tab:GGC2}) and in two cases
this number is larger than the estimated fluctuation of $N_{BD}$.
Given these numbers we cannot claim for any clear evidence of
lenses hosted in GGCs. Nonetheless, it is remarkable that for 
the 4 cases considered the value of $N_{GGC}$ is positive
and most probably underestimated, since the assumption that all the events 
lying in the outer ring are due to bulge/disc deflectors possibly overestimates $N_{BD}$.

\begin{figure}[here]
\begin{center}
\includegraphics[width=8 cm,height=6 cm]{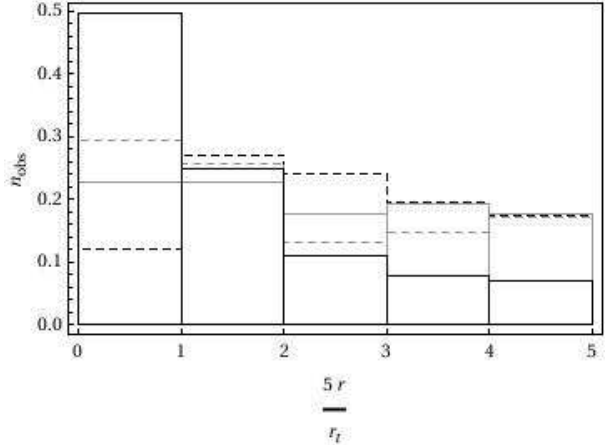}
\caption{ Barcharts of the number of detected events per unit area  as a function of distance from the center of the GGCs of Table \ref{tab:GGC2}. 
For every GGC of Table \ref{tab:GGC2}, we have divided its contour of radius $r=r_t$ in five circles 
centered at the GGC center and with radius $r_i = i \times r_t/5$, (i=1,..,5). In each of these circles 
we have calculated the number of detected events per unit area and have normalized the resulting barchart 
to unity (see text for more details). Black (solid,dashed) and gray (solid,dashed) lines show 
the barchart for NGC 6553, NGC 6522, NGC 6528 and NGC 6540, respectively.}

\label{fig:anelli}
\end{center}
\end{figure}

The fact that some events can be attributed to GGC MACHOs
can be further corroborated by the following consideration.
For a given GGC, we consider five circles of radius $r_i = i \times r_t/5$ (i=1,..,5) and centered at the GGC center. 
In every circle we compute the event density, $n_{obs,i}\equiv N_{obs,i}/A_i$, 
$N_{obs,i}$ being  the number of detected events inside the $i$-th circle and $A_i$ being the area of the $i$-th circle. 
For every GGC considered, the bar heights have been normalized to $\Sigma_{i=1}^5 n_{obs,i}=1$.

Fig. \ref{fig:anelli} shows the barcharts obtained for all the GGCs of Table \ref{tab:GGC2}.
In particular, for NGC 6553 (solid, black line) there is a clear excess of observed events toward 
its center. The same trend is found for NGC 6528 and NGC 6540 even though with less steep slopes, 
while NGC 6522 shows an inverse trend only in its innermost circle. Notice, anyways, that 
for all the
considered GGC between 40\% to 80\% of the events 
are located within $r=2\times r_t/5$ from the GGC center.

Assuming that the deflector is a GGC MACHO, we can estimate its mass
through the relation $R_E/t_E=v_r$, where $v_r$ is the 
lens-source relative velocity orthogonal to the l.o.s.,
$t_E$ is the event Einstein time and 
$R_E=\sqrt{\frac{4G}{c^2}m D_{ol}\big(1-\frac{D_{ol}}{D_{os}}\big)}$ 
is the Einstein radius, $m$ being the deflector mass.
As reported in \cite{harris}, the mean GGC tidal radius is of the order
of tens of pc,  this making the GGC extension relatively small compared to 
the average $D_{ol}$ and $D_{os}$ considered (of the order of kpc),
since we are assuming Galactic bulge/disc sources and the GGCs are kpcs 
away from the Sun.
For this reason, we make the simple assumption that in 
a given GGC the MACHOs are all at the 
same distance from the Sun ($r_{sun}$, as given in Table \ref{tab:GGC1}).
Furthermore, we assume all the sources to be on the Galactic $\{y,z\}$-plane \citep{straessle}.
As regards to $v_r$, we take advantage of recent results. In particular, except NGC 6540, all the GGCs listed in Table \ref{tab:GGC2} 
have been object of deep investigation and their proper motion has been described with considerable 
accuracy \citep{terndrup,zoccali,zoccalierratum,feltzing,dinescu}. 
In particular, in the reference frame $(U,V,W)$ where $U$ is positive outward from the GC toward the Sun, $V$ is positive toward the Galactic rotation
and $W$ is positive toward the Galactic Pole, we have $(27,57,-227)$ km/s for NGC 6522, $(-197,-26,4)$ km/s for NGC 6528 
and $(9,225,14)$ km/s for NGC 6553.
Moreover, due to the lack of information, 
we assume for NGC 6540 the reference value $v_r=100$ km/s 
and remind that for any other value, $\tilde v_r$, of the orthogonal relative velocity,
the corresponding lens mass, $\tilde m$, is simply rescaled as
$\tilde m = \big(\frac{\tilde v_r}{v_r}\big)^2 m$.
Table \ref{tab:GGC2} shows, for the whole subset of inner events, the predicted deflector mass in units of solar masses, $<m>$, 
obtained with these assumptions.
The resulting average lens mass gets values in the range $\{10^{-2},10\}$, suggesting that the involved deflectors are possibly 
either brown dwarfs, M-stars or stellar remnants. Moreover, Jupiter-like deflectors are not definitively excluded, since, 
already a small increase on $D_{os}$ can substantially reduce the predicted lens mass \footnote
{Following our analysis we find for both NGC 6522 and NGC 6528 an 
inner event whose predicted mass $m$ would exceed 10 $M_{\odot}$. 
These large values 
are essentially due to the term $\large[D_{ol}\large(1-D_{ol}/D_{os}\large)\large]^{-2}$ which takes into account
the lens geometry, that, with our choice on $D_{os}$ ($\sim 8.1~ kpc$)  
for both NGC 6522 and NGC 6528, becomes very small, since $D_{ol}$ is almost equal to $D_{os}$.
However, this last assumption might not hold as $D_{os}$ could be, for instance, larger, thus reducing the mass value substantially.
Indeed, already a 
slight variation by ($\sim2\%$) of $D_{os}$  can reduce the predicted lens mass by a factor 10.
Clearly, another possibility is that these two events are not due to lenses located in the GGC
and thus our mass estimate does not apply. 
Given these uncertainties we have not reported these two events in Fig. 3.}.

\begin{figure}[here]
\begin{center}
\includegraphics[width=8 cm,height=5 cm]{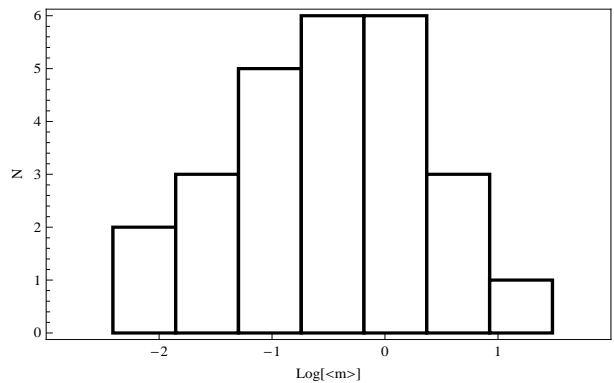}
\caption{Bar chart of the calculated lens mass, in units of solar masses, for the inner events.}
\label{fig:massa}
\end{center}
\end{figure}

The bar chart of the values of the expected lens mass
for the inner events is shown in Fig. \ref{fig:massa},
which highlights a crowding at $\sim 1\ M_{\odot}$.
The differences between the predicted $<m>$ that are given here
and the ones shown in \cite{straessle}
are due to the different Einstein time distributions, to the more appropriate GGCs proper motions adopted here
and to the fact that they assume $D_{os}=8.5\ kpc$ 
while we have $D_{os}\sim 8.1\ kpc$, with a very small dispersion among the targets.

The average expected lens mass has been drawn from the set of inner events,
some of which being possibly {\it not} due to GGC MACHOs.
This source of contamination should be removed before
one makes any prediction, but since we are not
able to do such a distinction, the average values on the overall inner sample can be
taken as a first crude approximation.

Also given in Table \ref{tab:GGC2} is the number of expected events toward
the GGC centers, $\Gamma_{exp}$, as calculated through formula (36) of \cite{straessle},
where it is assumed that all the lenses have the same mass, $\mu_o$,
in units of solar masses and that their distribution is
very narrow with respect to that of the source population. $\Gamma_{exp}$ is given in units of $f\times \mu_o^{-1/2}\times 10^{-3}/year$,
$f$ being the fraction of dark matter (in form of brown dwarfs, dim stars or stellar remnants) in the cluster.
For a typical value of $10^2-10^3$ monitored source stars behind a GGC (this number depending also on the GGC extension)
and an observation period of $\sim$ 5 to 10 years, we expect at most between half an event and a couple
of events toward each GGC depending also on the value of $f$, in reasonable
agreement with the results of Table \ref{tab:GGC2}.
Notice that $\Gamma_{exp}$ for NGC 6540 is almost 15 times larger than the one predicted for NGC 6528. This is due
to the larger central density of NGC 6540, which is a compact and massive GGC. 
On the other hand, since its extension is smaller, the number of 
aligned sources is quite small and thus the event rate
toward its center gets reduced.

\section{Conclusions} \label{sec:conclusions}	
We have analysed a large set of microlensing events detected toward the GC 
and have investigated the issue of whether and in which amount some of the
observed events can be attributed to MACHOs being hosted in GGCs.
We have first selected all the events
being aligned with a GGC 
and have taken into account the fact that still many 
of them indeed are lenses located in the Galactic bulge/disc.
We have estimated
the excess of events due to MACHOs in the GGC by dividing the area of each GGC
in two regions, an outer and an inner
one, and assuming that all the outer events are due to lenses not being hosted 
in the GGC and extrapolated the number of expected bulge/disc
events in the inner region.
We found that in a small region around 
the centers of four of the 135 overall considered 
GGCs some of the observed events could be attributed 
to lenses hosted by the GGC. 
Although all of the four cases, where there are a substantial amount
of events such as to be able to carry out our analysis, show an excess
of events, which could indicate the presence of lenses in the globular
clusters, we still cannot draw firm conclusions 
on this issue. 

Since the 
analysis made by \cite{straessle} the number of events
at disposal substantially
increased, but not yet to a level such that this question
can be answered in a definitive way. Clearly, the expected
number of events, and thus the rate, is certainly quite small
so that more observations are needed.
A possible strategy would be to survey systematically during many years
the line of sight comprising the four globular clusters
which we analysed.
Since we are not able to distinguish between the events due to the 
lenses in globular cluster
from the bulge/disc ones, for the estimate of the mass 
we just considered the mean time duration
of all events located in the inner region.
In this way we found a mass range suggesting that lenses in the GGCs
could possibly be brown dwarfs, M-stars or stellar remnants 
even though the uncertainties, especially on $D_{os}$, that affect our lens mass predictions
could also shrink the predicted mass interval and shift it toward the Jupiter-like objects domain.
In spite of all the above mentioned limitations,
we believe that our results, although not conclusive, suggest
that some events might indeed be due to lenses 
located in globular clusters.   
A dedicated survey 
over many years of the above considered globular clusters could
possibly resolve this issue.

F.D.L. acknowledges the Forschungskredit of 
the University of Zurich for financial 
support. The authors thank Mauro Sereno and Gaetano Scarpetta
for helpful suggestions and discussions.


\begin{thebibliography} {}

\expandafter\ifx\csname natexlab\endcsname\relax\def\natexlab#1{#1}\fi

\bibitem[{{Binney} \& {Tremaine}(1987)}]{binney}
{Binney}, J. \& {Tremaine}, S. 1987, {Galactic dynamics} (Princeton, NJ,
  Princeton University Press, 1987, p. 747)

\bibitem[{{Calchi Novati} {et~al.}(2007){Calchi Novati}, {De Luca}, {Jetzer},
  {Mancini}  \& {Scarpetta}}]{calchi07}
{Calchi Novati}, S., {De Luca}, F., {Jetzer}, P., {Mancini}, L. 
\& {Scarpetta}, G. 2007, to apper in A\&A, astro-ph 0711.3758

\bibitem[{{Cardone} \& {Cantiello}(2003)}]{cardone}{Cardone}, V. \& {Cantiello}, M. 2003, A\&A, 405, 125


\bibitem[{Dinescu {et~al.}(2003)}]{dinescu}{Dinescu}, D.~I. and {Girard}, T.~M. and {van Altena}, W.~F. \& {L{\'o}pez}, C.~E. 2003, AJ, 125, 1373

\bibitem[{{Feltzing} \& {Johnson} (2002)}]{feltzing}{Feltzing}, S. \& {Johnson}, R.~A. 2002, A\&A, 385, 67


\bibitem[{{Gyuk} \& {Holder}(1997)}]{gyuk}
{Gyuk}, G.. \& {Holder}, G.P. 1997, MNRAS, 297, L44


\bibitem[{{Hamadache} {et~al.}(2006){Hamadache}, {Le Guillou}, {Tisserand},
  {Afonso}, {Albert}, {Andersen}, {Ansari}, {Aubourg}, {Bareyre}, {Beaulieu},
  {Charlot}, {Coutures}, {Ferlet}, {Fouqu{\'e}}, {Glicenstein}, {Goldman},
  {Gould}, {Graff}, {Gros}, {Haissinski}, {de Kat}, {Lesquoy}, {Loup},
  {Magneville}, {Marquette}, {Maurice}, {Maury}, {Milsztajn}, {Moniez},
  {Palanque-Delabrouille}, {Perdereau}, {Rahal}, {Rich}, {Spiro},
  {Vidal-Madjar}, {Vigroux}, \& {Zylberajch}}]{eros2}
{Hamadache}, C., {Le Guillou}, L., {Tisserand}, P., {et~al.} 2006, A\&A, 454,
  185

\bibitem[{{Han} \& {Gould}(2003)}]{han&gould}
{Han}, C. \& {Gould}, A. 2003, ApJ, 592, 172

\bibitem[{{Harris}(1996) {Harris}}]{harris} {Harris}, W. E. 1996, AJ, 112, 1487


\bibitem[{{Heggie} \& {Hut} (1995)}]{heggie} {Heggie}, D. C. \& {Hut}, P. 1996, IUAS 174, 303H

\bibitem[{{Jetzer} {et~al.}(1998){Jetzer}, {Str\"assle}, \& {Wandeler}}]{straessle}
{Jetzer}, P., {Str\"assle}, M. \& {Wandeler}, U. 1998, A\&A, 336, 411


\bibitem[{{King} (1962) {King}}]{king} {King}, I. 1962, AJ, 67, 471

\bibitem[{{Kiraga} \& {Paczynski}(1994)}]{k&p}
{Kiraga}, M. \& {Paczynski}, B. 1994, ApJL, 430, L101


\bibitem[{{Rhoads} \& {Malhotra}(1998)}]{malhotra}
{Malhotra}, S. \& {Rhoads}, J.E. 1998, ApJ, 495, L55

\bibitem[{{Mandushev} {et~al.}(1991) {Mandushev} , {Staneva}, \& {Spasova}, N.}] {mandushev}
{Mandushev}, G., {Staneva}, A., \& {Spasova}, N. 1991, A\&A, 252, 94

\bibitem[{{Paczynski}(1986)}]{pacz86}
{Paczynski}, B. 1986, ApJ, 304, 1

\bibitem[{{Paczynski}(1994)}]{pacz94}
{Paczynski}, B. 1994, Acta Astronomica, 44, 235


\bibitem[{{Popowski} {et~al.}(2005){Popowski}, {Griest}, {Thomas}, {Cook},
  {Bennett}, {Becker}, {Alves}, {Minniti}, {Drake}, {Alcock}, {Allsman},
  {Axelrod}, {Freeman}, {Geha}, {Lehner}, {Marshall}, {Nelson}, {Peterson},
  {Quinn}, {Stubbs}, {Sutherland}, {Vandehei}, \& {Welch}}]{popo}
{Popowski}, P., {Griest}, K., {Thomas}, C.~L., {et~al.} 2005, ApJ, 631, 879


\bibitem[{{Sahu} {et~al.}(2001)}]{sahu}
{Sahu}, K.C., {Casertano}, L., {Livio}, M., {et~al.} 2001, Nature, 411, 1022


\bibitem[{{Stanek} {et~al.}(1997){Stanek}, {Udalski}, {Szymanski}, {Kaluzny},
  {Kubiak}, {Mateo}, \& {Krzeminski}}]{stanek}
{Stanek}, K.~Z., {Udalski}, A., {Szymanski}, M., {et~al.} 1997, ApJ, 477, 163

\bibitem[{{Sumi} {et~al.}(2006){Sumi}, {Wo{\'z}niak}, {Udalski},
  {Szyma{\'n}ski}, {Kubiak}, {Pietrzy{\'n}ski}, {Soszy{\'n}ski},
  {{\.Z}ebru{\'n}}, {Szewczyk}, {Wyrzykowski}, \& {Paczy{\'n}ski}}]{ogle}
{Sumi}, T., {Wo{\'z}niak}, P.~R., {Udalski}, A., {et~al.} 2006, ApJ, 636, 240

\bibitem[{{Taillet} {et~al.}(1995) {Taillet}, {Longaretti} \& {Salati}}]{taillet95} {Taillet}, R., {Longaretti}, P. Y. \& {Salati}, P. 1995, Astroparticle Physics 4, 87

\bibitem[{{Taillet} {et~al.}(1996) {Taillet},  {Salati} \& {Longaretti}}]{taillet96} {Taillet}, R.,  {Salati}, P. \& {Longaretti}, P. Y.  1996, ApJ, 461, 104 


\bibitem[{{Terndrup} {et~al.}(1998) {Terndrup}, {Popowski}, {Gould}, {Rich} \& {Sadler}}]{terndrup} 
        {Terndrup}, D.~M. and {Popowski}, P. and {Gould}, A. and {Rich}, R.~M. \& {Sadler}, E.~M. 1998, AJ, 115, 1476

\bibitem[{{Thomas} {et~al.}(2005){Thomas}, {Griest}, {Popowski}, {Cook},
  {Drake}, {Minniti}, {Myer}, {Alcock}, {Allsman}, {Alves}, {Axelrod},
  {Becker}, {Bennett}, {Freeman}, {Geha}, {Lehner}, {Marshall}, {Nelson},
  {Peterson}, {Quinn}, {Stubbs}, {Sutherland}, {Vandehei}, \& {Welch}}]{thomas}
{Thomas}, C.~L., {Griest}, K., {Popowski}, P., {et~al.} 2005, ApJ, 631, 906


\bibitem[{{Vallenari} {et~al.}(2006){Vallenari}, {Pasetto}, {Bertelli},
  {Chiosi}, {Spagna}, \& {Lattanzi}}]{vallenari}
{Vallenari}, A., {Pasetto}, S., {Bertelli}, G., {et~al.} 2006, A\&A, 451, 125

\bibitem[{{Wood} \& {Mao}(2005)}]{wood}
{Wood}, A. \& {Mao}, S. 2005, MNRAS, 362, 945


\bibitem[{{Zoccali} {et~al.}(2001)}]{zoccali}
{Zoccali}, M. and {Renzini}, A. and {Ortolani}, S. and {Bica}, E. \& {Barbuy}, B. 2001, AJ, 121, 2638

\bibitem[{{Zoccali} {et~al.}(2003)}]{zoccalierratum}
{Zoccali}, M. and {Renzini}, A. and {Ortolani}, S. and {Bica}, E. \& {Barbuy}, B. 2003, AJ, 125, 994





\end{thebibliography}
\end{document}